\documentclass[12pt]{article}

\usepackage{amsmath,amsfonts,amssymb,latexsym}
\usepackage{mathtools}
\usepackage{graphicx}
\usepackage{caption}
\usepackage{subcaption}

\usepackage[T1]{fontenc}
\usepackage[utf8]{inputenc}

\usepackage{microtype}

\usepackage[a4paper,left=2.2cm,right=2.2cm,top=2.6cm,bottom=2.6cm]{geometry}

\usepackage{setspace}
\PassOptionsToPackage{dvipsnames}{xcolor}
\usepackage{xcolor}
\usepackage{silence}
\usepackage{tikz}
\usepackage{tikz-3dplot}
\usetikzlibrary{arrows.meta,calc,decorations.markings}

\usepackage{amsthm}

\newtheorem{remark}{Remark}[section]

\numberwithin{equation}{section}

\usepackage{comment}

\def\be{\begin{equation}}
\def\ee{\end{equation}}
\def\bq{\begin{eqnarray}}
\def\eq{\end{eqnarray}}
\def\beq{\begin{eqnarray}}
\def\eeq{\end{eqnarray}}

\usepackage[colorlinks=true,
linkcolor=MidnightBlue,
citecolor=MidnightBlue,
urlcolor=MidnightBlue]{hyperref}
\begin{document}

%------------------------------------------------

\title{%
\textsc{Versal transition scenarios in inflationary cosmology: slow roll, ultra-slow roll, and oscillatory exit}
}

\author{\Large{\textsc{Spiros Cotsakis$^{1,2}$\thanks{\texttt{skot@aegean.gr}}}}\\
$^{1}$Clare Hall, University of Cambridge, \\
Herschel Road, Cambridge CB3 9AL, United Kingdom\\
$^{2}$Institute of Gravitation and Cosmology,  RUDN University\\
ul. Miklukho-Maklaya 6, Moscow 117198, Russia\\
}

\date{June  2026}

\maketitle

\begin{abstract}\noindent
We develop a physics-facing version of the persistence/transition-variety framework for scalar-field cosmology, tailored to inflationary dynamics. The guiding idea is that observationally viable inflationary models are often best understood not as single asymptotic phases but as concatenations of persistent regimes separated by universal transition episodes. In this picture, slow roll appears as a robust persistent balance, ultra-slow roll as a bottleneck passage near a nonhyperbolic organising set, and oscillatory post-inflationary behaviour as a recurrent exit sector. Using the exponential model as a reference regime atlas and the massive case as a dynamical realisation of slope drift, we show how such histories may be organised and read geometrically. The resulting framework makes explicit that the relevant regime transitions are organised precisely where hyperbolicity is lost or the spectrum crosses the imaginary axis, and are therefore invisible to a purely hyperbolic or asymptotic treatment.
\end{abstract}

\newpage
\tableofcontents
\newpage

\section{Introduction}
\label{sec:intro}

Inflationary models are often discussed in terms of the existence of a slow-roll regime and the predictions it yields for the basic observables~\cite{weinberg2,kl25}. More generally, the standard dynamical description of inflation is typically organised around asymptotic or quasi-asymptotic regimes --- vacuum-dominated expansion, quasi-de Sitter evolution, slow-roll attractors, or post-inflationary oscillatory behaviour --- viewed one phase at a time~\cite{bgzk85,linde1990,puzan}. This viewpoint has been highly successful, but it also encourages a reading of the dynamics in which the emphasis falls primarily on isolated regimes and their local stability properties~\cite{bgzk85,8}. Even within this standard picture, however, one is naturally led beyond the attractor itself whenever the relevant modes exit the horizon before full convergence to the attractor has occurred, so that the physically meaningful object is already the trajectory rather than the asymptotic state alone~\cite{puzan}.

Yet observational viability is increasingly controlled by the structure of the full inflationary history rather than by the mere presence of an inflationary attractor: one must also understand how a trajectory approaches the inflationary regime, whether it undergoes transient non-attractor or ultra-slow-roll episodes, how it exits inflation, and what oscillatory or reheating phase follows~\cite{MI-I}. This suggests a description in which the basic dynamical objects are not isolated asymptotic states alone, but concatenations of persistent regimes and universal transition events~\cite{persist}. In the present paper we develop such a description for scalar field cosmology, using the exponential model~\cite{cop1} as a reference stratification and the massive/quadratic case ~\cite{bgzk85} as the natural setting in which a drifting effective slope generates histories of the form slow roll \(\rightarrow\) ultra-slow roll \(\rightarrow\) oscillatory exit. Rather than reconstructing inflation retrospectively as a succession of named phases, the present framework organises the full evolution from the outset in terms of a regime atlas together with the transition geometry that links its components.

The present viewpoint shifts the emphasis from isolated phase-space objects to inflationary \emph{histories}, namely, orbits interpreted through the organising geometry of the stratification: the relevant dynamical content lies not in fixed points or phase curves alone, but in the way orbit segments, recurrent sets, and transition episodes are selected and linked by the underlying bifurcation structure. The approach adopted here combines two complementary strands. The first is the broader persistence framework for scalar-field cosmology, in which robust regimes are organised by stratification and linked by universal transition events~\cite{persist}. The second is the local inflationary analysis of the massive organising centre, where the reduced variables admit a direct interpretation in terms of slow variables and observables and where oscillatory sectors appear naturally in the lifted three-dimensional dynamics~\cite{MI-I,persist}. Our aim is to combine these two perspectives into a shorter and more physics-facing account tailored to inflationary cosmology. In the exponential family the slope parameter is fixed, so the stratification is read across the family as a regime atlas. In the harmonic or massive case, by contrast, the effective slope evolves along the orbit, so that the same stratification is sampled dynamically by a single history. This is precisely the mechanism by which concatenated histories are generated dynamically. In this sense, the paper is concerned less with isolated inflationary phases than with the architecture of complete inflationary histories. A systematic identification of broader classes of inflationary potentials sharing the same organising centres and unfolding structures — including plateau-type and trigonometric examples — lies beyond the scope of the present paper and will be discussed elsewhere. The present account is intentionally restricted to two transparent representatives, namely the fixed-slope exponential case and the dynamically varying-slope massive case.

What is new here, relative to the earlier and more technical treatments, is not a further extension of the general formalism as such, but its systematic inflationary reading. In particular, the present paper recasts slow roll, ultra-slow roll, and oscillatory exit as concatenated finite-time episodes organised within a single regime-atlas picture, and brings the role of nonhyperbolic organising loci to the foreground from the point of view of inflationary dynamics. The aim is therefore to make the persistence/unfolding framework directly readable in the language of inflationary histories and observables.

The paper is organised as follows. In Sec.~\ref{sec:regimemap} we introduce the
exponential scalar-field model as a reference stratification for inflationary
regimes and explain the physical meaning of its principal loci. In
Sec.~\ref{sec:episodes} we reinterpret slow roll, ultra-slow roll, and related
behaviours as persistent orbit segments connected by universal transition episodes.
In Sec.~\ref{sec:drift} we turn to the massive/quadratic case and show how a
drifting effective slope allows a single orbit to traverse several pulled-back
strata, thereby generating multi-stage inflationary histories dynamically. In
Sec.~\ref{sec:examples} we illustrate the framework with two worked
inflationary scenarios. Finally, in Sec.~\ref{sec:discussion} we discuss the role
of the present approach as a transportable classification scheme for inflationary
histories.

Throughout this paper we follow, as far as possible, the notation introduced in \cite{persist}. In particular, \(\gamma\) denotes the barotropic index of the background fluid, \(\lambda\) the logarithmic slope parameter associated with the scalar-field potential, and SR/USR abbreviate slow-roll and ultra-slow-roll, respectively. In the massive case we also use the reduced variables appropriate to the organising-centre analysis, so as to keep the link with the local inflationary description developed in \cite{MI-I}. This common notation allows the global stratification picture and the local reduced dynamics to be compared directly. Although the notation follows, as far as possible, that of \cite{persist}, the present paper is intended to be readable on its own. For this reason, the principal symbols and technical terms used below are recalled here or reintroduced at their first appearance, while only the more detailed structural material is deferred to \cite{persist}.

Because the present framework is based on persistence, organising loci, and versal unfolding ideas, some of the terminology used below is not standard in the inflation literature. For convenience, we therefore recall here the meanings of the principal terms as they are used in this paper.  By a \emph{stratum} we mean a connected piece of parameter or state space on which the qualitative regime structure is unchanged; a \emph{stratification} is the decomposition into such pieces. A \emph{robust regime} is a persistent dominant balance or mode of evolution (for example, a scalar-dominated slow-roll segment or a tracking balance), and a \emph{regime atlas} is the physical reading of the resulting stratification (for example, the \((\gamma,\lambda)\)-diagram of Fig.~1 when interpreted as a map of inflationary histories rather than as a purely formal bifurcation diagram). An \emph{organising locus} is a set at which this regime structure changes, typically through collisions, exchanges, or losses of hyperbolicity among relevant branches (in the simplest case, parameter-dependent equilibria). Here a \emph{collision} means that two relevant branches meet or merge at a distinguished parameter value, whereas an \emph{exchange} means that the dominant balance or dynamical role is transferred from one branch or regime to another across such a locus. A \emph{gate} is a distinguished invariant or symmetry-related set through which histories may pass, stall, switch branch, or be redirected, thereby mediating transitions between neighbouring regime segments. A \emph{hilltop-type structure} is a local bottleneck geometry near such a nonhyperbolic organising set, and a \emph{bottleneck episode} is the corresponding prolonged but finite passage of a history through that region. Finally, \emph{persistence} and \emph{non-persistence} refer respectively to the survival or loss of the qualitative regime under small perturbations or parameter variation. The later phrase \emph{pulled-back stratification} refers to the stratification induced along an actual history when a dynamical quantity such as the effective slope samples the fixed regime atlas. For standard background on bifurcation theory, normal forms, centre manifolds, and unfoldings, see, for example, \cite{gh83,GolubitskySchaefferI,ar94,wig03}.
\section{The inflationary regime map}
\label{sec:regimemap}
In this Section, we use the exponential scalar-field model as the reference stratification in the
\((\gamma,\lambda)\)-plane, and explain in a physics-facing way the principal strata,
their associated robust regimes, and the role of the main organising loci as
boundaries between qualitatively distinct inflationary behaviours. The emphasis
is on reading the map as a regime atlas for inflationary histories rather
than as a purely formal classification.

\begin{figure}[t]
\centering
\includegraphics[width=0.82\textwidth]{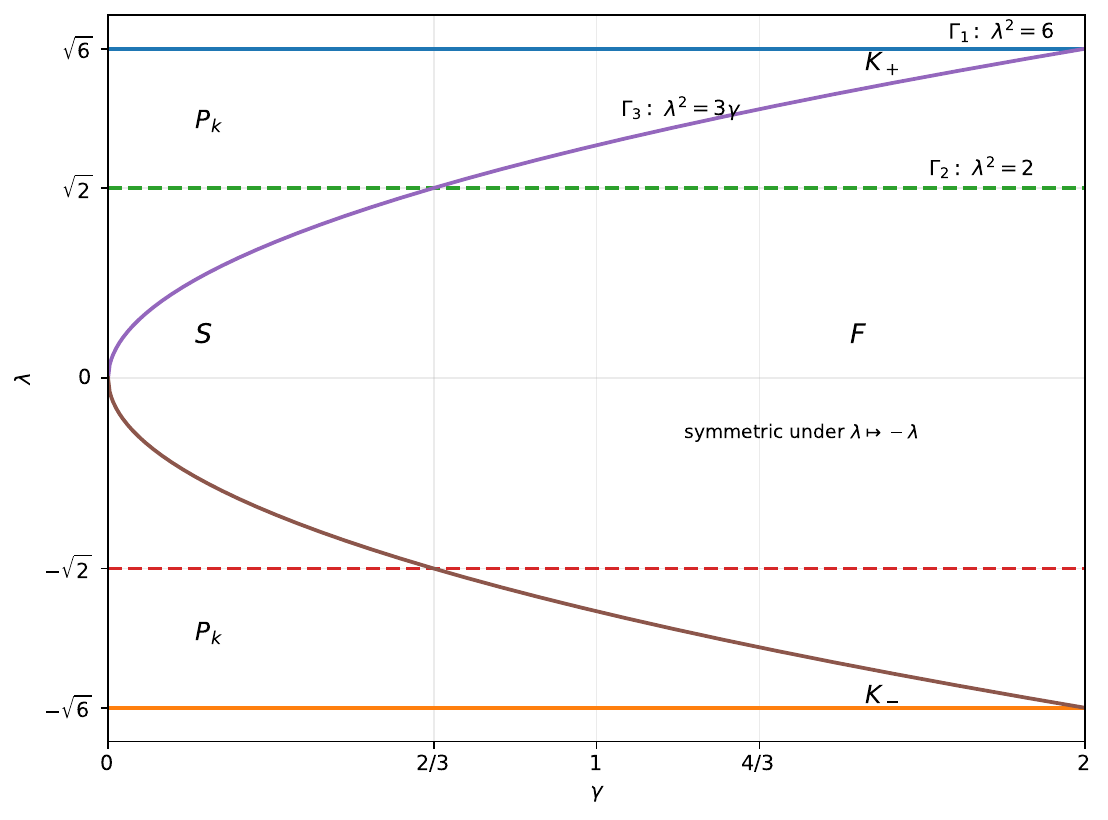}
\caption{Simplified inflationary regime map in the \((\gamma,\lambda)\)-plane for the exponential scalar--field family, showing only the three principal organising loci retained in the present paper: \(\Gamma_1\) (\(\lambda^2=6\)), \(\Gamma_2\) (\(\lambda^2=2\)), and \(\Gamma_3\) (\(\lambda^2=3\gamma\)). These separate the main sectors associated with the scalar-dominated branch \(S\), the curvature branch \(P_k\), the fluid-scaling branch \(F\), and the kinetic endpoints \(K_\pm\) relevant for the inflationary reading developed in the text. The picture is symmetric under \(\lambda\mapsto-\lambda\).}
\label{fig:regime-map}
\end{figure}

\subsection{Attractors vs. the regime atlas}
We begin with the exponential scalar-field model, not because realistic inflation must be exponential, but because it provides the cleanest fixed-slope template for reading the organisation of the full scalar--fluid--curvature system. In this setting the slope is a parameter rather than a dynamical variable, so the principal regime boundaries appear explicitly in the \((\gamma,\lambda)\)-plane and can be read as a genuine atlas of robust behaviours (cf. Fig.~\ref{fig:regime-map}). In this sense, the exponential family should be read here not merely as an isolated model choice, but as the universal fixed-slope representative of a wider class of inflationary potentials whose relevant regime is controlled by an exponential or asymptotically exponential germ.

This has a useful physical consequence for the present paper: curvature is not removed by assumption from the outset, but retained as a full mode, so that the distinction between scalar domination, curvature-sensitive evolution, tracking behaviour, and kinetic passages becomes part of the organising picture itself. 

The point of view adopted here is therefore slightly different from the standard one. Instead of asking only whether a given inflationary attractor exists, we ask how the relevant persistent regimes are arranged, where their boundaries lie, and what kinds of transition episodes occur when a history approaches those boundaries. In the strict exponential family this produces a fixed regime map. Later, in the massive or harmonic case, the same map will be sampled dynamically by a drifting effective slope. For the moment, however, the exponential model should be read simply as the reference geometry of the problem. 

At this point it is useful to recall that the full scalar--fluid--curvature system contains both parameter-independent equilibria and parameter-dependent branches \cite{persist}. The kinetic points \(K_\pm\), as well as the matter- and curvature-type equilibria \(M\) and \(C\), are present independently of the values of \((\gamma,\lambda)\), whereas the scalar-dominated, tracking, and curvature branches \(S\), \(F\), and \(P_k\) depend explicitly on the parameters and move accordingly through the state space. 

The principal loci \(\Gamma_i\) are not themselves equilibria, but organising sets in parameter space that record changes in the relation among these branches --- collisions, exchanges, or losses of robustness. In earlier discussions of inflationary dynamics the individual equilibria were of course known \cite{bgzk85,8,cop1}, but the organising role of these collision sets was not extracted systematically in the present stratification language; it is precisely this additional layer that turns the usual equilibrium picture into a genuine regime map \cite{persist}.

\subsection{The collision sets}
For the inflationary discussion it is enough to retain the three principal organising loci
\begin{equation}
\Gamma_1:\ \lambda^2=6,\qquad
\Gamma_2:\ \lambda^2=2,\qquad
\Gamma_3:\ \lambda^2=3\gamma,
\label{eq:three-main-loci}
\end{equation}
which separate the main robust sectors of the scalar--fluid--curvature system. In the notation inherited from \cite{persist}, these sectors are represented by the scalar-dominated branch \(S\), the fluid-scaling or tracking branch \(F\), the curvature branch \(P_k\), and the kinetic endpoints \(K_\pm\). The additional special loci discussed in \cite{persist} are certainly important in the full theory, but for the present inflation-centred account the geometry already becomes transparent at the level of \(\Gamma_1\), \(\Gamma_2\), and \(\Gamma_3\).

The first boundary, \(\Gamma_1\), is the outer kinetic threshold. It is the locus at which the scalar-dominated branch meets the kinetic endpoints, and therefore marks the edge of robust scalar domination. This is already visible from the scalar branch formula
\be
X_S=\sqrt{1-\frac{\lambda^2}{6}},\qquad
Y_S=\frac{\lambda}{\sqrt{6}},
\label{eq:scalar-branch}
\ee
since as \(\lambda^2\to 6\) one has \(X_S\to 0\) and \(Y_S\to \pm 1\), so that \(S\) collides with the kinetic endpoints \(K_\pm\). In physical terms, it separates histories that can settle into a scalar-supported regime from those in which kinetic behaviour takes over. Although \(\Gamma_1\) is not by itself an ``inflationary'' threshold, it is important for inflationary history because it organises kinetic passages and the onset of strongly non-slow-roll behaviour. 

In the broader persistence picture \(\Gamma_1\) is also the place where one first sees the local hilltop-type structure that later reappears as a prototype for bottleneck episodes \cite{persist}. A typical scalar-dominated example is a vacuum-driven inflationary stage in which the scalar field controls the dominant balance and the fluid and curvature sectors have become subdominant \cite{linde1990,puzan}. By contrast, a typical non-slow-roll example is a kinetic overshoot or kination-type passage, in which the potential no longer controls the evolution and the history is driven instead by the kinetic mode \cite{spokoiny93,peeblesvilenkin99}; for broader kinetic-sector inflationary mechanisms, see also \cite{kInflation}. In this sense \(\Gamma_1\) helps delimit not only the edge of scalar domination but also the onset of strongly non-slow-roll behaviour.

One conceptual advantage of the present framework is that familiar regime labels such as scalar domination acquire sharply defined dynamical boundaries. In particular, the edge of robust scalar domination is not a merely heuristic notion, but the explicit collision set \(\Gamma_1\), where the scalar branch \(S\) meets the kinetic endpoints \(K_\pm\).

The second boundary, \(\Gamma_2\), is the most immediately inflationary one. It is the scalar/curvature exchange line at which the scalar-dominated branch \(S\) meets the curvature branch \(P_k\). For the scalar solution itself, this is also the familiar borderline \(\lambda^2=2\) between accelerated and non-accelerated expansion. Indeed, along the scalar branch \(S\) one has
\begin{equation}
\epsilon = \frac{\lambda^2}{2},\qquad
w_\phi=\frac{\lambda^2}{3}-1,
\label{eq:scalar-branch-epsilon-wphi}
\end{equation}
where \(w_\phi:=p_\phi/\rho_\phi\) is the scalar-field equation-of-state parameter and \(\epsilon:=-\dot H/H^2\) is the usual Hubble slow-roll parameter. In particular, \(\epsilon<1\) corresponds to accelerated expansion, whereas \(\epsilon>1\) corresponds to decelerated expansion. Equivalently, for constant \(\epsilon\) one has \(a(t)\propto t^{1/\epsilon}\), so \(\ddot a>0\) for \(\epsilon<1\), \(\ddot a<0\) for \(\epsilon>1\), and the borderline \(\epsilon=1\) marks the transition between accelerated and decelerated expansion. Thus \(\Gamma_2\) is precisely the threshold \(\epsilon=1\) separating accelerated from non-accelerated scalar evolution. Equivalently, for constant \(\epsilon\) one has \(a(t)\propto t^{1/\epsilon}\), so \(\ddot a>0\) for \(\epsilon<1\), \(\ddot a<0\) for \(\epsilon>1\), and the borderline \(\epsilon=1\) marks the transition between accelerated and decelerated expansion. In this way, \(\Gamma_2\) is not merely a bifurcation-theoretic separator: it is the line at which the inflationary reading changes in a direct physical sense.

It is also useful to read this threshold through the scalar-field equation of state. At \(\lambda=0\) one has \(w_\phi=-1\), corresponding to the vacuum-energy limit, while at \(\Gamma_2\) one has \(w_\phi=-1/3\), so the effective gravitating combination \(\rho_\phi+3p_\phi=\rho_\phi(1+3w_\phi)\) vanishes. In this sense \(\Gamma_2\) is precisely the scalar-field threshold between decelerating and accelerating behaviour. By contrast, \(\Gamma_1\) gives \(w_\phi=1\), corresponding to the stiff or kinetic limit.

On one side of \(\Gamma_2\) one has a robust accelerated scalar regime; at the line itself curvature becomes dynamically competitive; beyond it the clean inflationary interpretation of scalar domination is lost. In this way the usual acceleration condition is embedded into a wider organising geometry rather than treated in isolation.

This is also the place where the inflationary significance of the collision becomes clearest. At \(\Gamma_2\) the scalar branch \(S\) and the curvature branch \(P_k\) meet in a nonhyperbolic way, and it is precisely through this collision that the accelerated scalar regime is organised. In that sense one may say that the collision is the geometrical event through which the inflationary scalar stage is organised. Such an event has no analogue in a purely hyperbolic classification, where branches persist without producing genuinely new accelerated sectors through collision.

The third boundary, \(\Gamma_3\), is the scalar/fluid exchange line. Here the scalar-dominated branch meets the tracking branch \(F\), and the system passes from a regime in which the scalar sector controls the late-time balance to one in which the scalar field scales with the ambient fluid. For inflationary purposes this boundary is valuable even when the fluid component is only auxiliary or effective, because it identifies the threshold at which scalar domination ceases to be structurally privileged. 

In other words, \(\Gamma_3\) marks the transition from inflation-like scalar control to tracking-type sharing of the dominant balance. This is one of the places where the full scalar--fluid--curvature formulation gives a richer picture than the scalar-only discussion usually adopted in the inflation literature.

\subsection{Persistence in the exponential class}
Taken together, the three loci in \eqref{eq:three-main-loci} divide the \((\gamma,\lambda)\)-plane into open regions as in Fig.~\ref{fig:regime-map} within which the qualitative regime structure is persistent. Each such region supports a definite dominant balance: scalar-dominated, curvature-sensitive, tracking, or kinetic. The important point is that these are not isolated solutions but robust sectors of behaviour. 

From the present viewpoint, an inflationary model is therefore read first by locating the region in which it lives and then by asking which boundaries its relevant histories approach. The regime map is static, but it already encodes a dynamical reading of possible histories. A simple example clarifies the distinction. In the exponential family, fixing \((\gamma,\lambda)\) selects a single point of the atlas and tells us once and for all which robust regime is relevant there; this is the static reading. 

In the massive case, by contrast, the effective slope varies along the history, so that one and the same solution may move across several pulled-back strata in succession --- for example, entering a slow-roll sector, lingering near a transition threshold, and later exiting toward a non-slow-roll or oscillatory phase. This is the dynamical reading.

This also clarifies the role of the strict exponential model in this paper. Because \(\lambda\) is fixed, the exponential family does not by itself generate multi-stage histories except by comparison across different members of the family. Its role is instead classificatory: it provides the reference regime atlas. The massive or harmonic case is more physically generative, because the effective slope becomes dynamical and a single orbit can pass through several pulled-back strata in succession. But without the exponential map one would not know what those passages mean. In this sense the fixed-slope picture is the static template against which dynamically generated inflationary histories are later read.

Accordingly, the message of the present section is simple. The exponential and massive scalar-field families play complementary roles in the present paper. The exponential case displays the fixed-slope regime atlas in its clearest form, whereas the massive case shows how that same organising structure can be traversed dynamically by a single history through the evolution of an effective slope. 

In the exponential atlas, the principal loci identify the thresholds at which scalar domination is lost, curvature becomes competitive, tracking sets in, or kinetic behaviour takes over. Once read in this way, the \((\gamma,\lambda)\)-plane becomes not just a bifurcation diagram but a regime map for inflationary histories. In the strict exponential family the relevant control parameters are external, so the map is sampled only by comparing different members of the family. 

Accordingly, the full set of organising loci is retained not because a single observationally viable exponential model traverses them, but because together they define the reference geometry against which more dynamical histories are later read.

In the later sections, however, unfolding parameters and slow variables will turn this static organisation into a genuinely dynamical one: paths through the stratification will then encode the transition episodes and concatenated histories themselves.

\section{Universal transition episodes}
\label{sec:episodes}
In this Section, we interpret slow roll, ultra-slow roll, exchange episodes, and kinetic passages as orbit segments linked by universal near-boundary transition events. We emphasize that the relevant dynamical objects are not isolated asymptotic
states alone, but persistent pieces of evolution together with the bottlenecks,
crossovers, and leakage events that connect them.

\subsection{From robust regimes to persistent pieces of history}

The regime atlas of Sec.~\ref{sec:regimemap} becomes physically meaningful only when read along actual histories. In this section, and throughout the paper, words such as \emph{stratum}, \emph{stratification}, \emph{organising locus}, \emph{episode}, and \emph{persistence} are used in the technical sense established in \cite{persist}, not merely as informal descriptive language. From the present viewpoint, slow roll, ultra-slow roll, exchange passages, and kinetic passages are not merely names for isolated asymptotic behaviours, but persistent, typically finite-time, pieces of evolution selected by the stratification and linked by universal near-boundary transition events. What matters dynamically is therefore not only which robust regime exists, but also how that regime is related to neighbouring ones, how a trajectory enters it, how long it dwells there, how it approaches an organising boundary, and how it leaves toward the next stage.

This differs in emphasis from the standard asymptotic reading of inflationary dynamics. In the more familiar picture, the central objects are attractors, fixed points, or late-time scaling states \cite{puzan,cop1}. Here such objects remain important, but they no longer exhaust the description. A finite but robust episode --- for example a prolonged passage near a nonhyperbolic threshold --- can be just as physically relevant as an asymptotic state. In this sense, the natural dynamical unit is not the attractor alone, but \emph{the regime segment together with the transition geometry that links it to neighbouring segments}.

A small mathematical example helps to make this distinction precise. Suppose that on an interval \(I=[N_1,N_2]\) of e-fold time the orbit remains inside a stratum where the same reduced dominant balance governs the motion, and that along this interval the slow-roll quantities satisfy
\be
\epsilon(N)\ll 1,\qquad |\eta(N)|\ll 1,\qquad N\in I.
\label{eq:sr-quantities}
\ee
Then the interval \(I\) is naturally interpreted as a slow-roll \emph{episode}: not because the solution has reached an infinite-time attractor, but because over that finite interval the same robust balance persists and controls the observable evolution. The mathematical content lies in the persistence of the governing balance on \(I\), not in the existence of a final asymptotic state at \(N\to\infty\).

A concrete example may help to fix the meaning of these terms. In the exponential class, the scalar-dominated branch \(S\) is given by \eqref{eq:scalar-branch} with $ \lambda^2<6$, in the notation of \cite{persist}. Along this branch one has
\be
w_\phi = 2Y_S^2-1=\frac{\lambda^2}{3}-1,
\qquad
\epsilon = 3Y_S^2=\frac{\lambda^2}{2}.
\label{eq:obs-scalar}
\ee
Hence the condition \(\lambda^2<2\) is exactly the condition \(\epsilon<1\) for accelerated scalar evolution. If an orbit remains for a finite interval of e-fold time in a neighbourhood of \(S\) with \(\lambda^2<2\), that interval is naturally interpreted as a slow-roll regime segment: the same dominant balance persists throughout it, \emph{even though the orbit need not converge to an asymptotic state}. For example, an orbit may spend \(50\)–\(60\) e-folds near the scalar branch \(S\), with \(\epsilon\ll 1\) and \(|\eta|\ll 1\), and then leave toward a bottleneck or exit phase. Such a history is inflationary for all practical and observational purposes, even if the orbit never converges asymptotically to the corresponding attractor. In this precise sense, the relevant object is a persistent finite-time episode rather than a final attractor alone.

The advantage of this viewpoint is that it treats inflationary evolution as a structured history rather than as a sequence of labels attached retrospectively to different portions of an orbit. A robust regime is read as a persistent segment of evolution: a stretch of history during which one dominant balance remains in control and the qualitative behaviour is stable under small perturbations. This need not mean that the orbit has reached an asymptotic end state or even that it remains in the same region indefinitely. It is enough that, over the interval of interest, the same organising balance governs the motion. In this sense, slow roll is naturally interpreted as a persistent attracting balance rather than merely as a fixed-point idealisation.

This point is especially important for inflationary phenomenology. The physically relevant observables are tied to a finite portion of the history, not to an infinite-time limit. What matters is often whether the trajectory has entered a slow-roll sector by the time the scales of interest exit the horizon, whether it remains there long enough, and whether it later approaches a threshold that produces a nontrivial transient episode \cite[Ch.~8, Sec.~8.2, and Eqs.~(8.34)–(8.35)]{mukhanov2005}, \cite{kinney2005}. The present framework is designed precisely for this situation: it does not discard the asymptotic picture, but refines it by admitting robust finite-time episodes as primary dynamical objects.

Accordingly, in the present framework inflationary regimes such as slow roll or ultra-slow roll are best understood not as asymptotic labels alone, but as persistent evolutionary episodes whose meaning is fixed by their position relative to the organising stratification. The role of the stratification is then twofold: it identifies the robust pieces of evolution, and it specifies the boundaries across which new episode types are created. In this sense, the distinctive contribution of the present framework is not to replace hyperbolic stability theory where it applies, but to describe the robust finite-time episodes and transition phenomena that arise precisely when the linear picture ceases to be decisive.

\subsection{Universal transition episodes near organising loci}

Once the dynamics is read in this way, the boundaries of the regime atlas acquire a direct physical meaning. They are not merely separators between labelled regions, but organising sets near which characteristic transition episodes occur. A trajectory that approaches one of these boundaries does not undergo an arbitrary change of behaviour. Rather, the local dynamics falls into recurring and structurally meaningful types: bottlenecks, exchanges, leakage events, or kinetic passages. It is these events that link one persistent piece of history to the next.

The clearest example is provided by a nonhyperbolic boundary passage (cf. Secs.~4 and~9.2 of \cite{persist}). Near such an organising locus the motion slows, the ordinary hyperbolic picture loses its descriptive sufficiency, and the history may spend a long but finite interval in a transitional state. This is exactly the kind of situation in which ultra-slow-roll behaviour should be understood in the present paper. It is not treated primarily as a separate asymptotic attractor, but as a bottleneck episode near a nonhyperbolic organising set: the trajectory lingers there, the physical evolution is strongly affected, and yet the episode is intrinsically transitional rather than final.

A simple normal-form model already shows why the finite-time language is preferable here. Near a generic bottleneck one is led to a reduced equation of fold type,
\begin{equation}
\dot u=\mu+u^2,\qquad 0<\mu\ll1,
\label{eq:bottleneck-normal-form}
\end{equation}
or to an equivalent one-dimensional reduction after a local change of variables. Here \(u\) is a local reduced coordinate along the slow passage direction near the nonhyperbolic organiser; it is introduced only as a normal-form variable and has no direct physical meaning by itself. The corresponding \emph{dwell time} means the characteristic time spent by the trajectory inside the bottleneck neighbourhood before exit. For example, if one measures the passage time across a fixed symmetric interval \([-U,U]\), one obtains
\begin{equation}
T(U;\mu)=\int_{-U}^{U}\frac{du}{\mu+u^2}
=\frac{2}{\sqrt{\mu}}\arctan\!\left(\frac{U}{\sqrt{\mu}}\right)
\sim \frac{\pi}{\sqrt{\mu}}
\qquad (\mu\to0^+),
\label{eq:bottleneck-time-scaling}
\end{equation}
and hence, up to an inessential constant factor,
\begin{equation}
T_{\mathrm{bottle}}\sim \mu^{-1/2}.
\label{eq:bottleneck-time-asymptotic}
\end{equation}
A concrete analogue appears in the quadratic organising picture of \cite{MI-I}, where the primary branches
\be
E_{1,2}=\bigl(\mp\sqrt{-\mu_1},0\bigr), \qquad \mu_1<0,
\ee
are created or annihilated through a saddle--node mechanism as the unfolding parameters cross the relevant boundary. In that setting the local bifurcation geometry is again not read as a terminal state alone, but as the organiser of the passage through a neighbourhood in which the accessible branches and their stability change. This makes the example from Ref.~\cite{MI-I} a useful prototype of how a long finite-time transition episode may arise from a local nonlinear organiser rather than from asymptotic convergence alone.

Thus the episode is finite, but it can be parametrically long. This is exactly the type of behaviour that is awkward in a purely asymptotic language and natural in the present one: the dynamical object is a long-lived transition episode, not a terminal state.

Other boundaries generate different but equally universal transition types. Near the scalar/fluid exchange line one encounters crossover behaviour in which the dominant balance is transferred from scalar domination to tracking-type evolution or conversely. Near the kinetic threshold one encounters passages in which the potential-supported regime loses control and kinetic behaviour becomes dynamically prominent. Near the scalar/curvature threshold one finds curvature leakage or recovery episodes, depending on the direction from which the boundary is approached. In each case the transition is best thought of not as an accident of one particular solution, but as the local realisation of a recurring organiser in the stratification.

A second brief illustration may be taken from the pitchfork-type birth of the branch \(Z_I\) in Case I of \cite{MI-I}. There the additional branch
\be 
Z_I=\left(-\frac{\mu_2}{n},\sqrt{-\frac{\mu_2^2}{n^2}-\mu_1}\right)
\ee
appears when the unfolding parameters cross the corresponding boundary, and the geometric interpretation is precisely that of a pitchfork-type transition in the reduced picture. The point for the present discussion is not the exact coefficients, but the fact that the change of dominant balance is encoded locally by a universal reduced geometry. The history does not merely jump from one label to another; it passes through a mathematically controlled exchange or birth of branches that reorganises the subsequent evolution.

The usefulness of this language is that it puts persistent regimes and transition episodes on the same footing. A complete inflationary history is no longer described merely by saying that it begins near one attractor and ends near another. Instead, it is read as a concatenation of robust segments connected by universal passage types. Symbolically, one should think not simply of
\[
\text{initial state} \longrightarrow \text{final state},
\]
but rather of
\[
\cdots \longrightarrow
\text{persistent regime} \longrightarrow
\text{transition episode} \longrightarrow
\text{persistent regime}
\longrightarrow \cdots .
\]
This is the sense in which the regime atlas of Sec.~\ref{sec:regimemap} becomes a tool for reading histories rather than merely classifying equilibria. The distinctive point of the present framework is therefore not merely that it uses nonlinear methods in a broad sense, but that it addresses inflationary regimes and transition episodes whose organising geometry is nonhyperbolic. In such situations the standard hyperbolic attractor picture, whether read locally through linearisation or globally through nonlinear stability arguments, is no longer sufficient by itself; the relevant persistence and transience are controlled instead by centre manifolds, normal forms, and unfoldings.

In the strict exponential family this concatenation picture is still read statically, by comparing where different members of the family sit relative to the organising loci. In the massive or harmonic case, however, the same transition types will later be realised dynamically along a single orbit as the effective slope drifts through the pulled-back stratification. The present subsection therefore supplies the conceptual bridge between the fixed regime atlas of the exponential model and the dynamically generated multi-stage histories considered below.

\subsection{Concatenation and the architecture of inflationary histories}

The previous discussion suggests a simple but important reformulation. An inflationary history should be read neither as a single asymptotic fate nor as a list of disconnected named phases, but as a concatenation of persistent finite-time regimes and universal transition episodes. In this reading, the basic question is no longer merely whether slow roll, ultra-slow roll, tracking, or kinetic behaviour can occur, but how these pieces may be linked and in what order.

This gives a more natural description of multi-stage histories. A typical evolution may contain a slow-roll segment, followed by a prolonged bottleneck episode, followed by a crossover into another regime, and later a kinetic or oscillatory exit. What is important is that these pieces are not assembled arbitrarily: the stratification constrains which transitions are possible, and the reduced local models determine how the corresponding episodes are realised. The geometry therefore organises not only the individual regimes, but also the admissible architecture of the histories themselves.

From this point of view, a symbolic sequence such as
\begin{equation}
\label{eq:history-symbolic}
\mathrm{SR}\longrightarrow \mathrm{USR}\longrightarrow \text{oscillatory exit}
\end{equation}
should not be read merely as a piece of descriptive shorthand. It stands for a definite dynamical scenario: a persistent slow-roll segment, followed by a bottleneck passage near a nonhyperbolic organiser, and then an exit into a regime whose oscillatory character is controlled by the next relevant organising structure. The usefulness of the notation is precisely that it compresses a geometrically organised history into a physically readable chain of episodes.

The asymptotic language is therefore not abandoned, but subordinated to a more flexible one. Attractors, sinks, scaling branches, and recurrent sets still identify important persistent pieces whenever they are present. What changes is that these are no longer taken to exhaust the physically relevant content of the history. Entry, dwell, transition, and exit become part of the primary dynamical description. In this sense, the concatenation viewpoint is not a rejection of the asymptotic picture, but its extension to the finite-time architecture of inflationary evolution.

This reformulation is especially useful for the next section. In the exponential class the concatenation picture remains largely schematic, since the slope is fixed and one compares different members of the family across the regime atlas. In the massive or harmonic case, however, the effective slope becomes dynamical, and the same symbolic chains are realised as actual paths through the pulled-back stratification. It is there that the history language becomes fully dynamical and that the geometrical notion of concatenation acquires its clearest physical force.

\section{Drifting slope and dynamically generated histories}
\label{sec:drift}
In this Section, we explain why the massive/quadratic case is the natural setting in which concatenated inflationary histories arise dynamically. We introduce only the minimum local and reduced-model material needed to show how the effective slope \(\lambda_{\mathrm{eff}}\) drifts, how a single orbit can traverse several pulled-back strata, and how histories of the form \eqref{eq:history-symbolic} emerge without external parameter retuning. The exponential and massive scalar-field families play complementary roles here: the former displays the regime atlas in fixed-slope form, while the latter shows how that same organising structure can be traversed dynamically by a single history.

\subsection{From fixed slope to drifting slope}

The central distinction in the present section is not between named potentials as such, but between two broader organising situations: fixed-slope and dynamically varying effective-slope dynamics. From the singularity-theoretic viewpoint, the pure exponential and the massive models are used as particularly transparent representatives of these two classes. What matters is not the full global form of the potential, but the germ and the unfolding that organise the relevant history.

For a canonical scalar field with potential \(V(\phi)\), we define the effective slope by
\begin{equation}
\label{eq:lambda-eff-def}
\lambda_{\mathrm{eff}}(\phi):=-\frac{V_{,\phi}}{V}.
\end{equation}
In the strict exponential family,
\begin{equation}
\label{eq:exp-potential}
V(\phi)=V_0 e^{-\lambda \phi},
\end{equation}
one has
\begin{equation}
\label{eq:lambda-exp-constant}
\lambda_{\mathrm{eff}}(\phi)\equiv \lambda=\mathrm{const.}
\end{equation}
Thus the \((\gamma,\lambda)\)-atlas of Sec.~\ref{sec:regimemap}, Fig.~\ref{fig:regime-map}, is read statically across the family: distinct choices of \(\lambda\) place distinct members of the family in different regions of the regime atlas, but a single orbit in a single model does not move across that atlas by any internal mechanism. Here, and in what follows, we use the word ``stratification'' in the broader regime-theoretic sense introduced in Sec.~\ref{sec:regimemap}; when the stricter bifurcation-theoretic meaning is intended, this will be clear from the explicit reference to unfolding parameters and phase-portrait equivalence.

By contrast, in the quadratic or massive case,
\begin{equation}
\label{eq:quadratic-potential}
V(\phi)=\frac12 m^2\phi^2,
\end{equation}
the effective slope is
\begin{equation}
\label{eq:lambda-eff-quadratic}
\lambda_{\mathrm{eff}}(\phi)=-\frac{2}{\phi},
\end{equation}
and therefore evolves along the orbit as \(\phi\) evolves. In particular, the slope is small for large \(|\phi|\), where the inflationary regime is realised, and grows in magnitude as the field approaches the minimum. This already shows in the simplest possible way that \emph{the same model may pass through different effective regions of the regime atlas during its evolution}.

A concrete dynamical form of this distinction is obtained by differentiating \eqref{eq:lambda-eff-def} with respect to e-fold time \(N=\ln a\). In the standard Hubble-normalised scalar-field variables one has
\begin{equation}
\label{eq:lambda-evolution-general}
\frac{d\lambda_{\mathrm{eff}}}{dN}
=
-\sqrt{6}\,\lambda_{\mathrm{eff}}^{\,2}\,(\Gamma_V-1)\,Y,
\qquad
\Gamma_V:=\frac{V\,V_{,\phi\phi}}{V_{,\phi}^{\,2}},
\end{equation}
where \(Y\) denotes the usual Hubble-normalised field-velocity variable. For the strict exponential potential, \(\Gamma_V\equiv 1\), and so \(\lambda_{\mathrm{eff}}' =0\), recovering \eqref{eq:lambda-exp-constant}. For the quadratic potential, \(\Gamma_V=\tfrac12\), and the slope necessarily drifts whenever \(Y\neq 0\). Thus the distinction between fixed-slope and drifting-slope histories is built directly into the reduced dynamics.

This point also answers a natural objection. Concatenated histories are not exclusive to the literal massive potential itself. More general exponential-like models, plateau models, or sums of exponentials may also exhibit dynamically varying effective slopes and hence dynamically generated concatenations. What is special about the strict exponential family is only that its slope is fixed, so that the regime atlas is displayed in its cleanest static form. The massive case is then the clearest representative of the complementary situation in which the same organising structure is traversed dynamically.

\subsection{Bounded-slope closure and invariant gates}

To make the drift picture geometrically transparent, it is convenient to replace the unbounded slope variable \(\lambda_{\mathrm{eff}}\) by the bounded angle variable already used in \cite{persist},
\begin{equation}
\label{eq:zeta-def}
\zeta:=\arctan \lambda_{\mathrm{eff}}.
\end{equation}
This compactifies the slope direction and allows the evolving effective slope to be read inside a bounded state space. In particular, the asymptotic sectors \(|\lambda_{\mathrm{eff}}|\to\infty\) are represented by the finite boundary values \(\zeta=\pm \frac{\pi}{2}\), while the flat-slope configuration \(\lambda_{\mathrm{eff}}=0\) becomes simply \(\zeta=0\).

Differentiating \eqref{eq:zeta-def} gives
\begin{equation}
\label{eq:zeta-evolution}
\frac{d\zeta}{dN}
=
\frac{1}{1+\lambda_{\mathrm{eff}}^{\,2}}
\frac{d\lambda_{\mathrm{eff}}}{dN}.
\end{equation}
Combining \eqref{eq:zeta-evolution} with \eqref{eq:lambda-evolution-general} shows immediately that the bounded-slope dynamics is regular even when \(|\lambda_{\mathrm{eff}}|\) becomes large. This is one of the reasons the \(\zeta\)-description is so useful: it turns the varying-slope problem into a compact geometric one. Indeed, substituting \eqref{eq:lambda-evolution-general} into \eqref{eq:zeta-evolution} gives
\begin{equation}
\label{eq:zeta-evolution-expanded}
\frac{d\zeta}{dN}
=
-\sqrt{6}\,
\frac{\lambda_{\mathrm{eff}}^{\,2}}{1+\lambda_{\mathrm{eff}}^{\,2}}
(\Gamma_V-1)Y,
\end{equation}
and the prefactor \(\lambda_{\mathrm{eff}}^{\,2}/(1+\lambda_{\mathrm{eff}}^{\,2})\) tends to \(1\) as \(|\lambda_{\mathrm{eff}}|\to\infty\), so the bounded-slope dynamics remains regular.

Within this bounded-slope closure, certain invariant sets and symmetry sets act as gates\footnote{Here \emph{gate} is used in the technical sense introduced in \cite{persist}: a distinguished invariant or symmetry-related set through which histories may pass, stall, switch branch, or be redirected, thereby mediating transitions between neighbouring regime segments. It is not meant as a merely colloquial label.} through which histories pass or at which their behaviour changes qualitatively. The two most important for the present discussion are the vanishing-field-velocity set
\begin{equation}
\label{eq:Y-gate}
Y=0,
\end{equation}
and the zero-slope set
\begin{equation}
\label{eq:zeta-gate}
\zeta=0.
\end{equation}
The first marks instants at which the scalar-field velocity vanishes and may subsequently change sign, that is, turning or reversal points of the scalar-field evolution. The second marks the passage through an effectively flat local slope, equivalently the locus \(\lambda_{\mathrm{eff}}=0\), where the potential has vanishing logarithmic slope. Thus \(Y=0\) is a dynamical gate in the velocity direction, while \(\zeta=0\) is a gate in the slope direction; histories crossing these sets may stall, reverse, or be redirected from one neighbouring regime segment to another (cf. section~7.2 of \cite{persist}).

A simple example is provided by the quadratic potential \eqref{eq:quadratic-potential}. Since \(\lambda_{\mathrm{eff}}=-2/\phi\), one has
\begin{equation}
\label{eq:zeta-quadratic}
\zeta=\arctan\!\left(-\frac{2}{\phi}\right),
\end{equation}
so that large-field inflationary motion corresponds to \(|\zeta|\ll 1\), whereas approach to the minimum drives \(\zeta\) toward the boundary values \(\pm\frac{\pi}{2}\). In this way the bounded variable \(\zeta\) records, in a compact and geometrically readable way, the passage from flat-slope inflationary behaviour to steep-slope exit behaviour.

The role of the bounded-slope system is therefore twofold. First, it regularises the slope evolution and makes the drift through different effective regimes mathematically manageable. Second, it identifies the local gates at which histories slow down, switch branch, or prepare to enter a new episode. In the next subsection these features are combined with the \((\gamma,\lambda)\)-atlas of Sec.~\ref{sec:regimemap} to produce a dynamically traversed version of the earlier static stratification.

\subsection{Pulled-back stratification and dynamically traversed histories}

The regime atlas of Sec.~\ref{sec:regimemap} is not discarded in the massive case. Rather, it is pulled back along the evolving effective slope. Symbolically, one replaces the fixed parameter value \(\lambda\) by the dynamical quantity \(\lambda_{\mathrm{eff}}(N)\), and the principal loci  \eqref{eq:three-main-loci} of Sec.~\ref{sec:regimemap}, 
%\begin{equation}
%\label{eq:gamma-loci-repeat}
%\Gamma_1:\lambda^2=6,
%\qquad
%\Gamma_2:\lambda^2=2,
%\qquad
%\Gamma_3:\lambda^2=3\gamma,
%\end{equation}
become dynamically traversed thresholds along a single history.

Equivalently, one may define the pulled-back transition sets
\begin{equation}
\label{eq:pulled-back-loci}
\widetilde{\Gamma}_1:=\{N:\lambda_{\mathrm{eff}}(N)^2=6\},
\qquad
\widetilde{\Gamma}_2:=\{N:\lambda_{\mathrm{eff}}(N)^2=2\},
\qquad
\widetilde{\Gamma}_3:=\{N:\lambda_{\mathrm{eff}}(N)^2=3\gamma\}.
\end{equation}
In the fixed-slope exponential family these are read only comparatively, by looking at different members of the family. In the varying-slope case they are encountered dynamically along one and the same orbit.

This is the sense in which the massive case turns the atlas into a \emph{generator of histories}. A single orbit may begin in a region where the effective slope is small, in the sense that \(|\lambda_{\mathrm{eff}}|\ll 1\), so that the potential is locally flat enough for the scalar field to dominate the background and support a slow-roll-type balance. It may then drift toward a threshold at which this scalar-dominated balance weakens, and then pass through a bottleneck, exchange, or exit episode organised by the local reduced dynamics. In this reading, a symbolic chain such as
\begin{equation}
\label{eq:history-chain-repeat}
\mathrm{SR}\longrightarrow \text{threshold passage}
\longrightarrow \text{new regime}
\end{equation}
is no longer merely schematic; it is realised by actual drift through the pulled-back stratification. In particular, one should think of the local reduced system not only as a catalogue of nearby invariant structures, but also as the organiser of actual traversing histories: trajectories of the full system may enter the neighbourhood of the organising centre, undergo a slowed or nonhyperbolic passage there, and then leave toward a different regime, thereby realising concrete pulled-back concatenations of the type described in the previous sections. A typical case to keep in mind is a monotone large-field history: it begins in a flat-slope slow-roll sector, drifts toward a nonhyperbolic threshold where a prolonged ultra-slow-roll-type bottleneck may occur, and then leaves toward a different regime, for instance the oscillatory neighbourhood of the minimum.

A useful way to phrase the difference is this. In the exponential class, the map is static and one learns from it by comparing models. For example, one may compare two exponential models with different fixed slopes, say one with \(\lambda^2<2\), which lies in the accelerated scalar sector, and another with \(\lambda^2>3\gamma\), which lies in a tracking-type sector. The comparison is informative because the two models occupy different regions of the same atlas, even though neither of them traverses those regions dynamically during its own evolution. In the massive class, by contrast, the same map is sampled dynamically by an evolving effective slope, and one learns from it precisely by following a single history through successive effective regions of the atlas. This is why the massive case is more physically generative for the present paper: it turns the regime language of Sects.~\ref{sec:regimemap}--\ref{sec:episodes} into an actual dynamical mechanism.

The key point is that the transition episodes of Sect.~\ref{sec:episodes} now arise without changing the model externally, that is, without retuning a fixed parameter from one member of the family to another.  The exponential class displays the fixed-slope regime atlas in its clearest form. The massive class shows how that same atlas is traversed dynamically, that is, crossed along an actual orbit through the evolution of \(\lambda_{\mathrm{eff}}\), and how its transition geometry becomes locally readable in terms of slow variables and observables. This is what makes the massive/quadratic case the natural bridge between the static regime atlas of the exponential class and the concatenated inflationary histories that are physically relevant.

\subsection{The local organising centre and the inflationary reading}

The global drift picture becomes especially powerful when combined with the local organising-centre analysis of Ref.~\cite{MI-I}. Near the massive organising centre, the reduced dynamics on the centre manifold is written in terms of the slow variables \((r,z)\) and takes the versal form
\begin{equation}
\label{eq:rz-versal-repeat}
\dot z=\mu_1+z^2+d\,r^2,
\qquad
\dot r=\mu_2 r+\frac32 r z+\cdots ,
\end{equation}
with \(d=\pm 1\) according to the organising side under consideration. Here the overdot denotes the local time variable of the reduced centre-manifold system. The variables \(r\) and \(z\) are local centre-manifold coordinates: \(r\) measures the size of the kinetic departure from the near-de Sitter balance, while \(z\) is the slow drift variable along the nonhyperbolic direction. The quantities \(\mu_1\) and \(\mu_2\) are the two unfolding parameters of the local versal family, and the sign \(d=\pm1\) distinguishes the organising side under consideration. The first equation governs the drift in \(z\), including the effect of the kinetic amplitude \(r\), whereas the second shows how \(r\) is damped or amplified through \(\mu_2\) and its coupling to \(z\). In this picture, the reduced variables are not merely technical auxiliaries: they acquire a direct inflationary interpretation. 

Their direct inflationary interpretation is given, to leading order, by the relations
\begin{equation}
\label{eq:eps-r}
\epsilon \sim \frac32 r^2,
\end{equation}
and
\begin{equation}
\label{eq:eta-z}
\eta \sim z,
\end{equation}
so that these relations are the local bridge between the organising-centre variables and the inflationary observables. 

This makes the regime language of the previous sections directly readable in local inflationary terms. A hyperbolic attracting sector with small \((r,z)\) is naturally read as a slow-roll segment. A prolonged passage near the nonhyperbolic organiser with \(\epsilon\ll 1\) but \(|\eta|=O(1)\) is naturally read as an ultra-slow-roll bottleneck episode. And on the oscillatory side, the emergence of periodic or quasi-periodic invariant structures gives the local model for post-inflationary oscillatory exit.

In this sense, the massive organising centre does more than provide another example. It is the point at which the two languages used in this paper meet. The global stratification language of \((\gamma,\lambda)\)-space, inherited from the exponential class, is converted into a dynamical history through the drift of \(\lambda_{\mathrm{eff}}\); the local versal language of \((r,z)\)-space then explains how that history is read in inflationary terms. This is why the massive/quadratic case is the natural setting in which concatenated histories become both dynamically realised and physically interpretable.

The conclusion of the present section is therefore simple. The exponential and massive scalar-field families are not competing models in the present analysis, but complementary representatives of two broader organising situations. The exponential class displays the fixed-slope regime atlas in its clearest form. The massive class shows how that same atlas is traversed dynamically, and how its transition geometry becomes locally readable in terms of slow variables and observables. It is this combination that turns the history language of the previous sections into a fully dynamical account of inflationary evolution.

The discussion of this section has therefore established the main dynamical mechanism of the paper. The regime atlas of the exponential class is not abandoned in the massive case, but reappears as a pulled-back organising structure traversed by a single evolving history. Through the drift of the effective slope and the local organising-centre geometry, concatenated inflationary sequences become dynamically realisable and physically interpretable. We now turn to explicit examples, first in the fixed-slope setting and then in the massive case, in order to show how the present framework reads concrete inflationary histories in practice.

\section{Worked inflationary scenarios}
\label{sec:examples}
In this Section, we illustrate the general framework on two representative examples. The first uses the strict exponential family as a fixed-slope prototype, showing how the regime atlas of Sec.~\ref{sec:regimemap} is read statically across different members of the family. The second turns to the harmonic or massive case, where the effective slope evolves along the orbit and the same organising structure is traversed dynamically by a single history. The aim is not to provide an exhaustive model analysis, but to show concretely how the language of regimes, transition episodes, and concatenated histories operates in practice.

\subsection{The fixed-slope baseline: reading the exponential atlas}

The exponential family provides the simplest setting in which the regime atlas can be read explicitly. Since the potential
\begin{equation}
\label{eq:exp-example-potential}
V(\phi)=V_0 e^{-\lambda\phi}
\end{equation}
has constant slope,
\begin{equation}
\label{eq:exp-example-slope}
\lambda_{\mathrm{eff}}(\phi)\equiv \lambda,
\end{equation}
each model corresponds to a single point in the \((\gamma,\lambda)\)-plane. In this sense, the exponential example is a reading exercise for the atlas rather than a dynamically generated concatenation: distinct members of the family occupy distinct sectors, but no single orbit crosses the principal loci by an internal mechanism.

For concreteness, fix a background fluid with
\begin{equation}
\label{eq:gamma-example}
\gamma=1.
\end{equation}
Then the principal thresholds of Sec.~\ref{sec:regimemap} occur at
\begin{equation}
\label{eq:thresholds-gamma-one}
\Gamma_2:\lambda^2=2,
\qquad
\Gamma_3:\lambda^2=3,
\qquad
\Gamma_1:\lambda^2=6.
\end{equation}
Three representative choices are then enough to display the basic logic of the atlas.

First, if
\begin{equation}
\label{eq:lambda-small-example}
\lambda^2<2,
\end{equation}
the model lies in the accelerated scalar sector. Along the scalar branch \(S\), one has
\begin{equation}
\label{eq:exp-scalar-branch}
X_S=\sqrt{1-\frac{\lambda^2}{6}},
\qquad
Y_S=\frac{\lambda}{\sqrt{6}},
\end{equation}
and hence
\begin{equation}
\label{eq:exp-epsilon}
\epsilon = 3Y_S^2=\frac{\lambda^2}{2},
\qquad
w_\phi = 2Y_S^2-1=\frac{\lambda^2}{3}-1.
\end{equation}
Thus \(\lambda^2<2\) implies \(\epsilon<1\), so the scalar branch supports accelerated expansion. A model in this sector is therefore read as belonging to a robust slow-roll-type regime, in the broad sense that the scalar-dominated balance is both persistent and inflationary.

Second, if
\begin{equation}
\label{eq:lambda-middle-example}
2<\lambda^2<3\gamma,
\end{equation}
with \(\gamma=1\) this means \(2<\lambda^2<3\), and the model remains scalar-dominated but no longer accelerated. The atlas then describes a clean loss of inflationary character without yet passing to the tracking branch. This is an example of how the regime map distinguishes between scalar domination as such and accelerated scalar domination: the former survives, the latter does not.

Third, if
\begin{equation}
\label{eq:lambda-large-example}
3\gamma<\lambda^2<6,
\end{equation}
the model passes to the tracking or fluid-scaling sector \(F\). The scalar field is no longer dynamically privileged in the same way, but instead shares the dominant balance with the background fluid. Finally, as \(\lambda^2\to 6\), the scalar branch collides with the kinetic endpoints \(K_\pm\), and the corresponding histories approach the kinetic threshold discussed in Sec.~\ref{sec:regimemap}.

The point of these examples is not to extract a sequence from a single model, but to show how the atlas is used. One learns from the exponential family by comparing models with different fixed values of \(\lambda\), and by identifying which region of the \((\gamma,\lambda)\)-plane each model occupies. The result is a static but physically informative classification: one sees immediately which models support accelerated scalar evolution, which lose acceleration while remaining scalar-dominated, and which enter tracking or kinetic sectors. In this sense the exponential example is the fixed-slope reference case against which the dynamically generated histories of the massive case should be read.

It is important to stress, however, that for a literal exponential inflationary model compatible with current observations, the relevant regime lies deep in the small-\(\lambda\) accelerated scalar sector. In that sense, the loci \(\Gamma_1\) and \(\Gamma_3\) are not introduced here because one observationally viable fixed-slope exponential model is expected to approach them dynamically, but because the exponential family is being used as a reference regime atlas. Their role in the present paper is therefore organising and comparative rather than directly phenomenological within a single fixed exponential model.

\subsection{A dynamically generated history in the massive case}

The massive example differs conceptually from the exponential one in a decisive way. For the quadratic potential
\begin{equation}
\label{eq:massive-example-potential}
V(\phi)=\frac12 m^2\phi^2,
\end{equation}
the effective slope is
\begin{equation}
\label{eq:massive-example-slope}
\lambda_{\mathrm{eff}}(\phi)=-\frac{V_{,\phi}}{V}=-\frac{2}{\phi},
\end{equation}
and therefore varies along the orbit. In the bounded notation of Sec.~\ref{sec:drift}, this becomes
\begin{equation}
\label{eq:massive-example-zeta}
\zeta=\arctan\lambda_{\mathrm{eff}}
=\arctan\!\left(-\frac{2}{\phi}\right).
\end{equation}
Thus the same model samples different effective values of the slope as the field evolves, and a single history can traverse the pulled-back stratification dynamically.

This immediately yields a natural qualitative scenario. At large field values one has
\begin{equation}
\label{eq:large-field-small-slope}
|\phi|\gg 1
\quad\Longrightarrow\quad
|\lambda_{\mathrm{eff}}|\ll 1,
\end{equation}
so the effective slope is small and the potential is locally flat enough to support a scalar-dominated slow-roll segment. As the orbit evolves and \(|\phi|\) decreases, the slope grows in magnitude,
\begin{equation}
\label{eq:slope-growth}
|\phi|\downarrow
\quad\Longrightarrow\quad
|\lambda_{\mathrm{eff}}|\uparrow,
\end{equation}
and the history is driven toward the relevant pulled-back thresholds
\begin{equation}
\label{eq:pulled-back-thresholds-repeat}
\widetilde{\Gamma}_1:\lambda_{\mathrm{eff}}^2=6,
\qquad
\widetilde{\Gamma}_2:\lambda_{\mathrm{eff}}^2=2,
\qquad
\widetilde{\Gamma}_3:\lambda_{\mathrm{eff}}^2=3\gamma.
\end{equation}
A typical monotone history should be imagined as follows: the orbit begins at large field values in a flat-slope slow-roll sector, then drifts toward a nonhyperbolic threshold where a prolonged ultra-slow-roll-type bottleneck may occur, and finally leaves toward the oscillatory neighbourhood of the minimum. In this way, the qualitative trajectory itself realises the symbolic sequence of regimes.
The regime atlas of Sec.~\ref{sec:regimemap} is therefore no longer merely a catalogue of distinct models: it becomes a sequence of thresholds encountered by one evolving solution.

In this way the symbolic history
\begin{equation}
\label{eq:massive-history-symbolic}
\mathrm{SR}\longrightarrow \mathrm{USR}\longrightarrow \text{oscillatory exit}
\end{equation}
acquires a direct dynamical meaning. The first segment corresponds to the large-field, small-slope phase in which the scalar balance is robust and inflationary. The middle segment corresponds to a slowed passage near a nonhyperbolic organising structure, where the ordinary hyperbolic attractor picture becomes insufficient and a bottleneck or ultra-slow-roll-type episode may occur. The final segment corresponds to the approach to the minimum, where the scalar dynamics becomes oscillatory and the inflationary phase terminates.

This should not be read as a claim that every quadratic trajectory exhibits an identically realised sequence in exactly the same way. Rather, it shows that the massive case contains the natural mechanism by which such concatenations can arise without external parameter retuning. The history itself provides the traversal parameter. In particular, the drift of \(\lambda_{\mathrm{eff}}\) means that the transitions discussed abstractly in Sec.~\ref{sec:episodes} are now realised along an actual orbit.

A useful summary is therefore the following. In the exponential family the slope is fixed and one compares models. In the massive family the slope evolves and one follows a single history. The former displays the regime architecture statically; the latter generates it dynamically.

\subsection{What the examples show}

The two examples considered above should be read together. The exponential family supplies the fixed-slope baseline and teaches the reader how to use the regime atlas of Sec.~\ref{sec:regimemap}. The massive family then shows how that same atlas is converted into an actual history by the internal evolution of the effective slope. The two are therefore complementary representatives, not competing model choices.

This also clarifies why the balance tilts toward the massive example in the present paper. In the exponential class one reads the geometry by comparing models placed in different sectors of the atlas. In the massive class one reads the same geometry by following a single history through successive effective regions. The former is classificatory; the latter is dynamically generative.

In this sense the exponential model provides the reference atlas, whereas the harmonic or massive case provides the physically decisive example: the former displays the regime architecture statically, while the latter generates it dynamically through the drift of the effective slope. What the examples therefore show is not simply that two familiar potentials behave differently, but that a history-based organisation of inflation naturally separates fixed-slope and dynamically varying effective-slope situations.

The broader lesson is methodological. Once the relevant objects are understood as germs and unfoldings rather than global potentials taken one by one, the role of the examples changes. The pure exponential and massive models are used here because they are especially transparent representatives of two organising classes. Their value lies in the fact that they make the history architecture mathematically visible. In this way the examples illustrate the central message of the paper: inflationary models are most naturally compared not by isolated attractors alone, but by the regime segments and transition episodes that organise their histories.

\section{Discussion: towards a classification of inflationary histories}
\label{sec:discussion}

The main purpose of this paper has been to argue that inflationary dynamics is more naturally organised by histories than by isolated attractors alone. In the framework developed here, the relevant dynamical objects are persistent finite-time regime segments and the universal transition episodes that link them. Slow roll, ultra-slow roll, exchange passages, and oscillatory exit are therefore read not merely as named asymptotic phases, but as geometrically organised pieces of a larger inflationary history.

A second purpose of the paper has been to provide a shorter and more physics-facing route into the broader persistence framework developed in \cite{persist}, by isolating the inflationary regime atlas, the associated transition episodes, and the dynamical mechanism by which these are realised in concrete histories. In this way the paper is intended not as a replacement for the fuller framework, but as an entry point into it, centred on the problems most directly relevant to inflationary cosmology.

The conceptual gain of this viewpoint may be stated succinctly. The exponential and massive scalar-field families are used here as complementary representatives of two broader organising situations. The exponential case displays the fixed-slope regime atlas in its clearest form, while the massive case shows how that same organising structure can be traversed dynamically by a single history through the evolution of an effective slope. The examples of Sec.~\ref{sec:examples} then show how the symbolic chain
\[
\mathrm{SR}\longrightarrow \mathrm{USR}\longrightarrow \text{oscillatory exit}
\]
is not merely descriptive shorthand, but a mathematically controlled concatenation of persistent regimes and transition episodes.

This perspective also clarifies the status of the asymptotic picture. The present framework does not deny the usefulness of attractors, sinks, scaling branches, or recurrent sets wherever these exist. Rather, it places them inside a broader language adapted to finite-time architecture. In many inflationary applications, the physically relevant observables are determined during a finite window of the evolution, before any strict asymptotic limit is reached. For that reason, a persistent finite-time slow-roll or ultra-slow-roll episode may be more physically relevant than the infinite-time fate of the orbit, and the transition geometry leading into and out of such an episode becomes part of the primary dynamical description.

From the mathematical side, the phenomena treated here are inseparable from nonhyperbolicity. The transitions that organise the histories discussed in this paper occur precisely where branches collide, hyperbolicity is lost, or the relevant spectrum crosses the imaginary axis. They are therefore invisible to any treatment that remains confined to a single bounded hyperbolic region or that relies exclusively on asymptotic phase-portrait analysis within one fixed stratum. In this sense, the regime transitions, bottlenecks, and concatenations that form the main target of the present paper are genuinely nonhyperbolic objects, and their proper description requires centre manifolds, normal forms, and unfoldings.

This last point is especially important for the inflationary interpretation. In standard treatments, the existence of a slow-roll attractor often provides the organising background for the discussion. Here, by contrast, the decisive structures are often those that arise when the hyperbolic picture ceases to be sufficient: bottleneck passages near nonhyperbolic organising sets, exchange episodes across collision loci, and locally oscillatory sectors generated by the unfolding of the massive organising centre. These are not corrections to an otherwise complete hyperbolic story; they are part of the geometry needed to understand inflationary histories in their full dynamical form.

One conceptual advantage of the present analysis is that it replaces qualitative regime language by sharply defined dynamical boundaries. Scalar domination, accelerated scalar evolution, tracking, and kinetic behaviour are no longer separated by informal crossover descriptions alone, but by explicit organising loci at which branches collide, exchange, or cease to persist. This is why collisions such as those at \(\Gamma_2\) can be said to organise an inflationary stage geometrically, something with no direct analogue in a purely hyperbolic classification.

A further possible implication concerns the classical initial-data problem for inflation. The present framework does not determine the ultimate origin of the initial state, nor does it replace questions about quantum cosmology or the microscopic preparation of the inflaton \cite{kolbturner}. It does, however, suggest a weaker and more flexible dynamical requirement than the usual attractor-based one: what matters physically is not only whether an orbit converges asymptotically to a slow-roll state, but whether it enters and persists for a sufficiently long finite interval in the relevant regime segment or near the appropriate nonhyperbolic organiser. In this sense, the set of dynamically acceptable histories may be broader than in a purely asymptotic reading, and the dependence on a specially prepared pre-inflationary state may be correspondingly softened. This also suggests that thermal preparation of the inflaton should not be regarded as conceptually privileged in the present setting: the relevant dynamical requirement is entry into an appropriate persistent or bottleneck-controlled history, not an explanation of primordial specialness through thermalization itself. In this sense, the present framework is consistent with the view that thermalization by itself does not explain the specialness of the early universe: what matters here is not a thermal origin of inflationary initial data as such, but the dynamical accessibility of the appropriate history architecture once the system enters the relevant regime \cite{penrose2004}.

The broader methodological lesson is that inflationary models should not be compared solely as isolated potentials taken one by one, but through the germs and unfoldings that organise their histories. From this viewpoint the pure exponential and massive models are not privileged because of their names, but because they are particularly transparent representatives of fixed-slope and dynamically varying effective-slope situations. This suggests a wider program, namely the classification of broader families of inflationary potentials by their organising centres and transition structures rather than by global potential shape alone.

Such a classification lies beyond the scope of the present paper, but the route toward it is already visible. The examples considered here suggest that many observationally viable inflationary models may be compared more naturally through a small number of regime architectures than through a proliferation of superficially distinct potentials. In this sense, the present work may be read as a first step toward a transportable classification of inflationary histories: one in which entry, dwell, transition, and exit are treated as the basic dynamical data, while the singularity-theoretic structure remains in the background as the organising principle that makes those histories mathematically legible.

\addcontentsline{toc}{section}{Acknowledgments}
\section*{Acknowledgments}
I thank Prof. G. Leontaris for his kind invitation to contribute this paper to the special issue of \emph{Universe} in celebration of Prof. I. Antoniadis’ 70th birthday. Ignatios has been a close friend and valued collaborator to me for more than two decades. I am deeply grateful to him for his unique insights and open-mindedness, and I wish him many more years of fruitful contributions to theoretical physics.


\addcontentsline{toc}{section}{References}

\begin{thebibliography}{99}
\bibitem{weinberg2}
S. W. Weinberg, \emph{Cosmology} (OUP, Oxford, 2007)
\bibitem{kl25} 
R. Kallosh, A. Linde, \emph{On the present status of inflationary cosmology}, Gen. Rel. Grav. 57 (2025) 10, 135; arXiv:2505.13646 [hep-th].

\bibitem{bgzk85} 
V. A. Belinski, L. P. Grishchuk, Ya. B. Zel'dovich, and I. M. Khalatnikov, \emph{Inflationary stages in cosmological models with a scalar field}, Sov. Phys. JETP \textbf{62} (1985) 195, Fig.~1.

\bibitem{linde1990}
A.~Linde,
\emph{Particle Physics and Inflationary Cosmology}
(CRC Press, London, UK, 1990), Sec.~1.6.

\bibitem{puzan}
P.~Peter and J.-P.~Uzan,
\emph{Primordial Cosmology}
(Oxford University Press, 2009), Ch.~8.

\bibitem{8} G. W. Gibbons, S. W. Hawking, and J. M. Stewart, \emph{A Natural Measure on the Set of All Universes,} Nucl. Phys. B \textbf{281} (1987) 736.

\bibitem{MI-I}
S. Cotsakis and I. Antoniadis, \emph{Mode interactions in scalar field cosmology}, Phil. Trans. R. Soc. (2026) (accepted); arXiv:2512.04607v2.
\bibitem{persist}
S. Cotsakis, \emph{Persistence and Transition Varieties in Scalar Field Cosmology};  arXiv:2604.05617.

\bibitem{cop1}
E. J. Copeland, A. R. Liddle, and D. Wands,
\emph{Exponential potentials and cosmological scaling solutions},
Phys. Rev. D \textbf{57} (1998) 4686.

\bibitem{gh83}J. Guckenheimer and P. Holmes, \emph{Nonlinear oscillations, dynamical systems, and bifurcations of vector fields} (Springer, 1983)
\bibitem{GolubitskySchaefferI}M. Golubitsky,  D. G. Schaeffer, \emph{Singularities and Groups in Bifurcation Theory,} Volume I (Springer, 1988)
\bibitem{ar94}V. I. Arnold, \emph{Dynamical Systems V: Bifurcation Theory and Catastrophe Theory }(Springer, 1994)
\bibitem{wig03}S. Wiggins, \emph{Introduction to applied nonlinear dynamical systems and chaos,} 2nd. Ed. (Springer, 2003)



\bibitem{spokoiny93}
B. Spokoiny,
\emph{Deflationary universe scenario},
Phys. Lett. B \textbf{315} (1993) 40.

\bibitem{peeblesvilenkin99}
P. J. E. Peebles and A. Vilenkin,
\emph{Quintessential inflation},
Phys. Rev. D \textbf{59} (1999) 063505.

\bibitem{kInflation}
C. Armend\'ariz-Pic\'on, T. Damour, and V. Mukhanov,
\emph{k-Inflation},
Phys. Lett. B \textbf{458} (1999) 209--218; arXiv:hep-th/9904075.

\bibitem{mukhanov2005}
V. Mukhanov,
\emph{Physical Foundations of Cosmology}
(Cambridge University Press, Cambridge, 2005).

\bibitem{kinney2005}
W. H. Kinney,
\emph{Horizon crossing and inflation with large eta},
Phys. Rev. D \textbf{72} (2005) 023515.

\bibitem{kolbturner}
E. W. Kolb and M. S. Turner, \emph{The Early Universe}
(Addison-Wesley, Redwood City, CA, 1990), Sec.~8.2, pp.~269, 317.

\bibitem{penrose2004}
R. Penrose, \emph{The Road to Reality: A Complete Guide to the Laws of the Universe}
(Vintage Books, London, 2004), Sec.~28.5, p.~755.


\end{thebibliography}
\end{document}